\documentclass{webofc}
\usepackage[varg]{txfonts}   % Web of Conferences font
\usepackage{cleveref}
\usepackage{mathtools}
\usepackage{xcolor}
\begin{document}
\title{Inverse problems, real-time dynamics and lattice simulations}

\author{\firstname{Alexander} \lastname{Rothkopf}\inst{1,3}\fnsep\thanks{\email{alexander.rothkopf@uis.no}}
        % etc.
}

\institute{Faculty of Science and Technology, University of Stavanger, NO-4021 Stavanger, Norway}

\abstract{%
  The determination of real-time dynamics of strongly coupled quantum fields is a central goal of modern nuclear and particle physics, which requires insight into quantum field theory beyond the weak-coupling approximation. While lattice QCD has provided vital insights into the non-perturbative static properties of quarks and gluons it hides their real-time dynamics behind an ill-posed inverse problem. In this proceeding I will discuss developments in tackling the inverse problem on the lattice and touch upon progress in the direct simualtion of real-time dynamics. 
}
\maketitle
\section{Introduction}

The task of determining the real-time dynamics of strongly correlated quantum fields arises in many of the topical tracks of this conference, be it in the study of the transport properties of the hot quark-gluon plasma created in relativistic heavy-ion collisions in \texttt{Track D} or the relaxation of the dense interior of collisions of two neutron stars in \texttt{Track F}. With the electron-ion collider project underway the phase-space tomography of nucleons and eventually nuclei constitutes a central topic in \texttt{Track A} and \texttt{Track B}. Similarly the precision determination of particle properties, such as the muon $g-2$ requires high precision real-time input from the theory of the strong interactions, quantum chromodynamics (QCD), as discussed in contributions to \texttt{Track E}.

Lattice QCD over the past decade has proven to be a robust tool to elucidate strong interaction physics (see e.g. \cite{Ratti:2018ksb} or \cite{Davoudi:2022bnl}). One sets out to evaluate expectation values of observables in Feynman's path integral from first-principles on a hypercubic grid with lattice spacing $a$
\begin{align}
\langle {\cal O}(x_1){\cal O}(x_2)\rangle=\int {\cal D}[U,\bar\psi,\psi]\,{\cal O}[x_1;U,\bar\psi,\psi]{\cal O}[x_2;U,\bar\psi,\psi]\,{\rm exp}\Big[iS_{\rm QCD}[U,\bar\psi,\psi]\Big].
\end{align}
In order to preserve the gauge invariance of the theory after discretization, the $su(3)$ valued gauge fields are formulated in terms of so called link variables $U_\mu(x)={\rm exp}[-igA^\mu(x+a\hat \mu/2)]$ located in between the nodes of the grid. Quark fields $\psi,\bar\psi$ on the other hand are distributed on the nodes. 

In its Minkowski time form the Feynman weight ${\rm exp}[-i S]$ amounts to a pure phase, rendering the evaluation of this path integral by means of naive statistical sampling exponentially hard. The success underlying lattice QCD simulations is the realization that we may analytically continue the real-time path integral onto the negative imaginary time axis, i.e. into the Euclidean domain. There the Feynman weight ${\rm exp}[-S_E]$ is real-valued and bounded from below, which offers a statistical interpretation of the path integral and makes its evaluation amenable to established sampling approaches, such as Markov-Chain Monte-Carlo. I.e one generates a collection of field realizations distributed according to the Feynman weight, computes the value of operators on each of these, and by taking the mean systematically approximates the quantum-statistical expectation value of interest. Due to the subtle but vital relation between the physical extent of the Euclidean domain and the inverse temperature of the system, introducing periodic boundary condition in imaginary time is not a discretization artifact but actually encodes a vital physical property of the system.

Our original intention is to evaluate Minkowski-time correlation functions, which in Euclidean time on the lattice are not directly accessible. One central player in the quest of extracting such real-time information is the spectral function. The spectral function $\rho(\omega)$ of a two-point function refers to the unique quantity, which, through an integral transform with a known kernel function $K$, encodes all of the information of the correlator in a representation independent fashion, i.e. irrespective of whether the two-point function is expressed as retarded or advanced real-time quantity or on the Euclidean domain
\begin{align}
{\cal O}^E(\tau)=\int d\omega\,K(\omega,\tau)\rho(\omega) \quad \overset{\rm analytic\, continuation}{\leftrightarrow} \quad \int d\omega\,K(\omega,it\pm i\varepsilon)\rho(\omega)={\cal O}^{R/A}(t)\label{eq:specdec}.
\end{align}
Importantly the spectral function itself often encodes in an intuitive fashion the physics of interest. Take as an example the heavy quarkonium spectral functions studied in \texttt{Track C} and \texttt{Track D}. The low frequency behavior of $\rho(\omega)$ encodes transport properties, such as heavy quark diffusion via linear response theory (Kubo formula) and at intermediate frequencies the in-medium hadron bound-state properties are encoded in peak structures (for a more detailed discussion see e.g. \cite{Rothkopf:2019ipj}).

\section{Spectral function reconstruction from Euclidean lattices}

\begin{figure}
\centering
\includegraphics[scale=0.45]{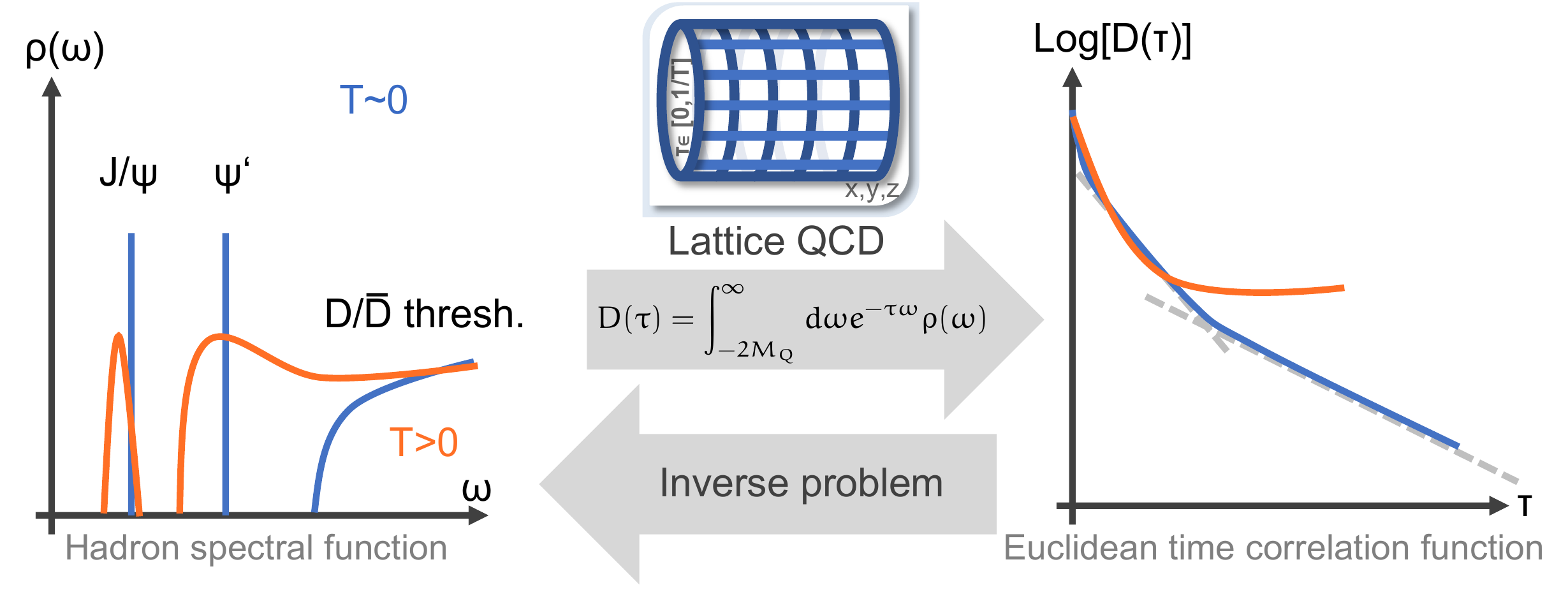}
\caption{Sketch of the inverse problem using quarkonium as example. We must extract spectral functions $\rho$ from Euclidean lattice simulations, which only produce a noisy and discrete imaginary time correlator $D$ that corresponds to a convolution of the spectral function with a known kernel function K.} \label{fig:InvProb}\vspace{-0.75 cm}
\end{figure}

We thus face the task of extracting spectral functions from Euclidean simulations. It turns out that lattice QCD acts on spectral functions very similarly to an imperfect detector acts on particle tracks. The integral transform of \cref{eq:specdec} smears out the intricate structures present in the spectral function (see \cref{fig:InvProb}) into an imaginary time correlator $D$, based on which we must solve an inverse problem to regain access to the values of $\rho$. The challenge lies in the fact that a Monte-Carlo simulation produces only discrete and noisy realizations of the correlation functions and that the kernel in the convolution can lead to exponential loss of information when going from $\rho$ to $D$.

In some circumstances the inverse problem to reconstruct $\rho$ is benign, when the spectral structures are well separated and their form is known apriori. In that case, e.g. when extracting the grounds state masses of hadrons at $T=0$, a parametrized model of the spectral structure can be constructed which is directly fitted to the correlator $D$. In other circumstances, e.g. when extracting in-medium hadron properties, intricate features appear in the spectral function that are not well separated and no simple closed form parametrization of $\rho$ is available. In that case the inverse problem is clearly ill-posed in the sense of Hadamard: there exist many different realizations of $\rho$ which reproduce the noisy correlator $D$ within its statistical uncertainties (non-uniqueness) and due to the exponential behavior of the kernel $K$ the inverse problem without additional regularization is highly sensitive to noise in the input data (instability).

The exponential information loss induced by the integral kernel $K$ is a common issue when reconstructing spectral functions in different contexts. It is obvious from the functional form of $K$ when considering hadron physics at $T=0$ and $T>0$
\begin{align}
K^{T=0}={\rm exp}[-\omega \tau], \quad K^{T>0}={\rm cosh}[\omega(\tau-\beta/2)]/{\rm sinh}[\omega\beta/2], \quad K^{\rm PDF}_{\rm Ioffe}={\rm cos}[x\nu]
\end{align}
However one faces an additional challenge, due to limited reach of the simulations. At finite temperature, we are apriori limited to accessing Euclidean times only up to $\tau<\beta$, which leads to the inherent coarseness of Matsubara frequencies. It is this coarseness which may prevent the successful reconstruction of essential spectral features related to thermal physics. In the study of PDFs in the pseudo-PDF formalism \cite{HadStruc:2021qdf} one is in practice limited to resolving only a subset of Ioffe-time values, which leads to exponentially small eigenvalues in the kernel matrix $K$ associated with the otherwise innocent looking kernel ${\cos}[\nu x]$. Keeping these challenges in mind, we must ask how to extract most accurately the information residing inside the simulated correlators.

Three strategies can be found in the literature: \texttt{Strategy I} does not accept the premise that lattice QCD can provide us with "the spectral function". Instead it settles for extracting a smeared approximation of the spectral function from a linear combination of the input data. This strategy is implemented using the modified Backus-Gilbert method \cite{backus1968resolving,hansen2019extraction}. \texttt{Strategy II} accepts the premise and attempts to model the spectral structures using prior domain information input, derived e.g. from low-energy effective field theories or perturbation theory. This leads to a sophisticated ansatz for the spectral function fitted to the correlator (see e.g. \cite{mages2015shear,burnier2017thermal,Astrakhantsev:2018oue}). \texttt{Strategy III} also accepts the premise but refrains from a direct parametrization. It includes prior information on the level of a regulator functional introduced in the context of Bayes theorem. The Bayesian inference approach encompasses various methods from sparse modelling, Tikhonov regularization \cite{tikhonov1943stability} and even some neural network approaches.

In the following I will focus on Bayesian inference \cite{jarrell1996bayesian}, a framework which is also applied in various contexts in the contributions to \texttt{Track H}. The discrete spectral relation we wish to invert reads
\begin{align}
D_i = \sum_{k=1}^{N_\omega} \, \Delta \omega_k K(\omega_k,\tau_i)\,\rho(\omega_i) \quad i\in[1,\ldots,N_\tau],\;N_\omega\gg N_\tau.
\end{align}
I.e. we wish to reconstruct the $N_\omega$ parameters $\rho_k=\rho(\omega_k)$ from $N_\tau$ noisy datapoints. To give meaning to this task, one deploys Bayes theorem, which describes the conditional probability for a test function $\rho$ to be the correct spectral function underlying supplied simulation data $D$ and prior information $I$. This so called posterior probability is rewritten as a product
\begin{align}
\underbracket{P[\rho|D,I]}_{\rm posterior}= \underbracket{P[D,\rho,I]}_{\rm likelihood}\underbracket{P[\rho|I]}_{\rm prior}/\underbracket{P[D|I]}_{\rm evidence} \quad \overset{\rm MAP}{\rightarrow} \quad \left.\delta P[\rho|D,I]/\delta\rho \right|_{\rho=\rho^{\rm MAP}}=0
\end{align}
of the likelihood probability $P[D,\rho,I]\propto{\rm exp}[-L]$, which encodes information about the data generation process and the prior probability $P[\rho|I]\propto{\rm exp}[S]$, which can be used to encode domain knowledge beyond what is provided by the simulation data. Normalization is provided by the so-called evidence.
In order to save computational cost, one often does not sample to full posterior but instead computes the most likely spectral function $\rho^{\rm MAP}$, the maximum aposteriori, which denotes the extremum of the posterior. In the absence of prior information one ends up with many degenerate maximum likelihood solutions. The functional $S$ inside the prior is therefore called \textit{regulator}, as it picks from among the many degenerate local extrema of the likelihood the one that aligns closest with given domain information (e.g. positivity of a hadronic spectral function).

The prior probability is conventionally characterized by its two first cumulants, the so-called default model $m_l$ and (hyperparameter) weight $\alpha_l$. Whereas $m_l$ encodes the most probable apriori value of the of the parameter $\rho_l$ its apriori variance is described by $1/\alpha_l$. All Bayesian approaches by construction must agree in the \textit{Bayesian continuum limit} of simultaneously increasing the number of data points $N_\tau\to\infty$ and reducing their uncertainty $\Delta D/D\to0$. The choice of $S$ influences which types of artifacts one encounters on the way, such as e.g. ringing for weak regulators or overdamping of spectral structures for strong regulators.

Its versatility allows the Bayesian approach to encompass several regularization schemes. The well-known Tikhonov approach is found to correspond to a Gaussian prior probability with a vanishing default model and a weight $\alpha$ set to a value that recovers $L=N_\tau/2$ for $\rho^{\rm MAP}$. Note that this approach does not exploit the positivity of hadronic spectral functions as its regulators admits also negative values for the $\rho_l$'s. Two other Bayesian approaches, which are commonly deployed in the study of positive definite spectral functions are the Maximum Entropy Method (MEM) \cite{jarrell1996bayesian,asakawa2001maximum} and the Bayesian Reconstruction (BR) \cite{burnier2013bayesian} method, which feature the following regulators
\begin{align}
S_{MEM}=\alpha\int d\omega\Big(\rho-m-\rho{\rm log}\Big[\frac{\rho}{m}\Big]\Big), \quad S_{BR}=\alpha\int d\omega\Big(1-\frac{\rho}{m}+{\rm log}\Big[\frac{\rho}{m}\Big]\Big).
\end{align}
The former was originally constructed for the reconstruction of images in astronomy and is based on four axioms. For a detailed discussion of the axioms see \cite{skilling1988axioms}. The hyperparameter $\alpha$ is treated in a self consistent fashion in the literature by finding the extremum of the evidence, a procedure that relies on a saddle-point approximation of the posterior, which is otherwise analytically intractable.

The BR method is also based on four axioms \cite{burnier2013bayesian}, which are selected with the one-dimensional reconstruction problem in lattice QCD in mind. One of the axioms is scale invariance, which in essence requires that the end result of the reconstruction may not depend on the units of the spectral function and is takes care of the fact that $\rho$ is not necessarily a probability distribution itself. The second axiom to note is smoothness, which requires that the end result should be a smooth (at least twice differentiable) function. The functional form of $S_{BR}$ allows for a semi-analytic apriori marginalization of $\alpha$, which offers a truly Bayesian treatment of this often poorly constrained hyperparameter.

Bayesian spectral function reconstruction has been deployed to the study of $T>0$ physics and $T=0$ physics from the lattice. The study of in-medium hadron spectral functions, in particular those of heavy-quarkonium bound states \cite{Aarts:2014cda,Kim:2018yhk} is a central application, with new results presented by the FASTSUM collaboration in \texttt{Track D} at this conference (for a recent study of excited states quarkonium see \cite{Larsen:2019zqv}). Progress has been made by using Bayesian spectral reconstructions in the study of the complex in-medium potential between static quarks from lattice QCD \cite{Burnier:2015tda,Burnier:2016mxc}. Recent results on the potential from lattices with dynamical quarks \cite{Bala:2021fkm} will be presented by Gaurang Parkar in \texttt{Track C}. Over the past years first steps have been taken to extract parton distribution functions from the lattice using among others Bayesian inference \cite{Liang:2019frk,Karpie:2019eiq}, an overview over the state-of-the-art in PDFs from the lattice is presented in \texttt{Track B}.

The past years have also seen heightened interest in machine learning approaches to tackle the spectral function reconstruction challenge. In the literature one finds direct neural network approaches, where methods used in image reconstruction are transferred to spectral functions. The situation here is often actually simpler than in image reconstruction since the decoder step of going from the spectral function to the data is analytically known. Two recent approaches are supervised Kernel-Ridge regression \cite{Offler:2021fmg} or support vector machine regression \cite{Kades:2019wtd,fournier2020artificial}, which differ by the basis functions used, the loss functional they deploy and the regulators included to obtain unique solution. 

A different approach to regularization has been explored in recent papers \cite{Karpie:2019eiq,Shi:2022yqw} where a neural network is used in the parametrization of the spectral function itself $\rho(\omega)={\rm NN}(\omega)$. This parametrization, which in essence amounts to a certain choice of basis functions can be used in both supervised and unsupervised settings. It is important to note that the regularization of the reconstruction is still implemented through additional terms in the loss functional that guides the training of these neural networks. 

A very recent study \cite{Horak:2021syv} has turned to another well established machine-learning technique, Gaussian processes, to reconstruct spectral functions. It models both observed data and predicted data as arising from the same underlying Gaussian distribution. It is the correlation matrix of this high dimensional normal distribution that implements the regularization of the approach. The hyperparameters of the correlation matrix are self-consistently tuned based on training data.

If we take a step back and ask, where do we stand in regard to spectral function reconstruction we can recognize that so far no clear methods winner has emerged. The central limitation remains the scarce information content of the correlator and one pressing open question is: how do we include more QCD specific prior knowledge into the regularization of the inversion task? It is important to note that significant increases in statistics by now enable the use of complimentary methods, such as e.g. the Pad\'e reconstruction \cite{Tripolt:2018xeo}. 

As the community works on improving lattice simulations to more efficiently approach the continuum limit (Szymanzik improvement) one encounters competing interests: while improved actions may accelerate the approach to the continuum, they can interfere with conventional spectral decomposition by introducing e.g. negative spectral weights (see discussion in \cite{Bala:2021fkm}). A better understanding of these artifacts may allow us to incorporate and potentially correct for them in improved regulator terms.

Given the challenges faced in spectral function reconstruction from Euclidean lattices, one may ask whether direct simulation strategies for real-time dynamics can provide complementary insight.

\section{Direct real-time simulations}

The notorious sign problem \cite{gattringer_approaches_2016} prevents a straight forward evaluation of the Minkowski-time path integral using established Monte-Carlo approaches. On the one hand we may find scenarios, where the quantum fluctuations of fields are small compared to the statistical fluctuations. This approximation underlies the classical statistical lattice gauge theory (CLGT) framework, which has been successfully used to shed light on the non-equilibrium physics of the early stages of heavy-ion collisions (see e.g. \cite{Berges:2013fga}), as well as on topology changing processes in thermal equilibrium (see e.g. \cite{Moore:2010jd}). In a classical statistical simulation, the gauge fields are evolved deterministically according to the classical equations of motion, initialized by values drawn from a thermal distribution. Carrying out the deterministic evolution for multiple different stochastic initial values allows us to form ensemble averages of observables at finite real-time.

On the other hand one may attempt to design alternative techniques for the direct evaluation of the real-time path integral. The two most promising strategies in our community are the Lefschetz thimbles \cite{cristoforetti_new_2012} and complex Langevin \cite{namiki_stochastic_1992}, both of which rely on a complexification of the field degrees of freedom, moving the path integration into the complex plane. The challenge with Lefschetz thimbles lies in identifying the corresponding submanifolds, by solving a flow equation, which can incur large numerical cost in practice. For complex Langevin, the main drawback lies in the fact that while the approach may converge, it does not necessarily converge to the correct solution.

Taking up a line of inquiry started in \cite{Laine:2007qy}, we recently used classical statistical simulations to explore the binding of static quark-antiquark pairs in real-time \cite{Lehmann:2020fjt}. For the first time it was possible to compute the real part of the interaction potential by incorporating the physics of the sources into the Gauss law of the simulation. The static potential is related to the time evolution of the Wilson loop, which describes the propagation of a pair of a static quark and antiquark. It has been shown on general grounds \cite{Burnier:2012az} that if the Wilson loop evolves according to a Schr\"odinger equation at late times, its spectral function, which in Minkowski time is obtained from a simple Fourier transform of the real-time Wilson loop, will exhibit a dominant skewed Lorentzian peak
\begin{align}
\nonumber i\partial_t \langle W_\square(r,t)\rangle\sim\Big({\rm Re}&[V](r)-i|{\rm Im}[V](r)|\Big)\langle W_\square(r,t)\rangle ,\qquad \overset{\rm Fourier}{\leftrightarrow} \\
&\langle W_\square(r,\omega)\rangle \sim c_0\frac{{\rm Im}[V]^2(r)}{(\omega-{\rm Re}[V](r))^2+{\rm Im}[V]^2(r)} + {\rm background}.
\end{align}
I.e. the position and width of the spectral peak encodes the real- and imaginary part of the complex potential. Note that the evaluation of the Wilson loop in a fully quantum path integral or as an observable in a classical statistical simulation differs substantially. In the former setting the evaluation of the Wilson loop actually corresponds to a reweighting of a system without static quarks to a system where the sources are present,  i.e. $\langle W(r,\tau)\rangle_{\rm path\,integral}=Z_{\rm medium\,+\,Q\bar{Q}}(r,\tau)/Z_{\rm medium}$. in classical statistical simulations on the other hand the Wilson loop is simply evaluated as observable and does not affect the real-time dynamics of the gauge fields. Sources instead need to be treated as explicit part of the equations of motion, where their charge density $\sigma$ enters as
\begin{align}
Z_{\rm CLGT}&=\int {\cal D}A(t=0)\int {\cal D}\Pi(t=0)\rho(A,\Pi,t=0)\delta\Big(D_\mu F^{\mu\nu}[A]-j^\nu\Big),\\
j^0&=M\Big(\delta^{(3)}({\bf x}-{\bf x}_0)-\delta^{(3)}({\bf x}-{\bf x}_1)\Big)={\color{red} \sigma({\bf x};{\bf x}_0,{\bf x}_1)}.
\end{align}
I.e. we must modify the local Gauss law on the lattice by hand, when introducing static sources in the classical statistical simulation
\begin{align}
G({\bf x},t)\equiv\sum_i\Big[ E_i({\bf x},t)-U_i({\bf x},t)E_i({\bf x}+{\bf i},t)U_{-i}^\dagger({\bf x},t)\Big]={\color{red} \sigma({\bf x};{\bf x}_0,{\bf x}_1)}.
\end{align}
This version of Gauss' law amended by the static charge density enters the classical statistical dynamics solely via the initial conditions, which must be projected onto the constraint surface given by $G({\bf x},t)={\color{red} \sigma({\bf x};{\bf x}_0,{\bf x}_1)}$.

\begin{figure}
\includegraphics[scale=0.35]{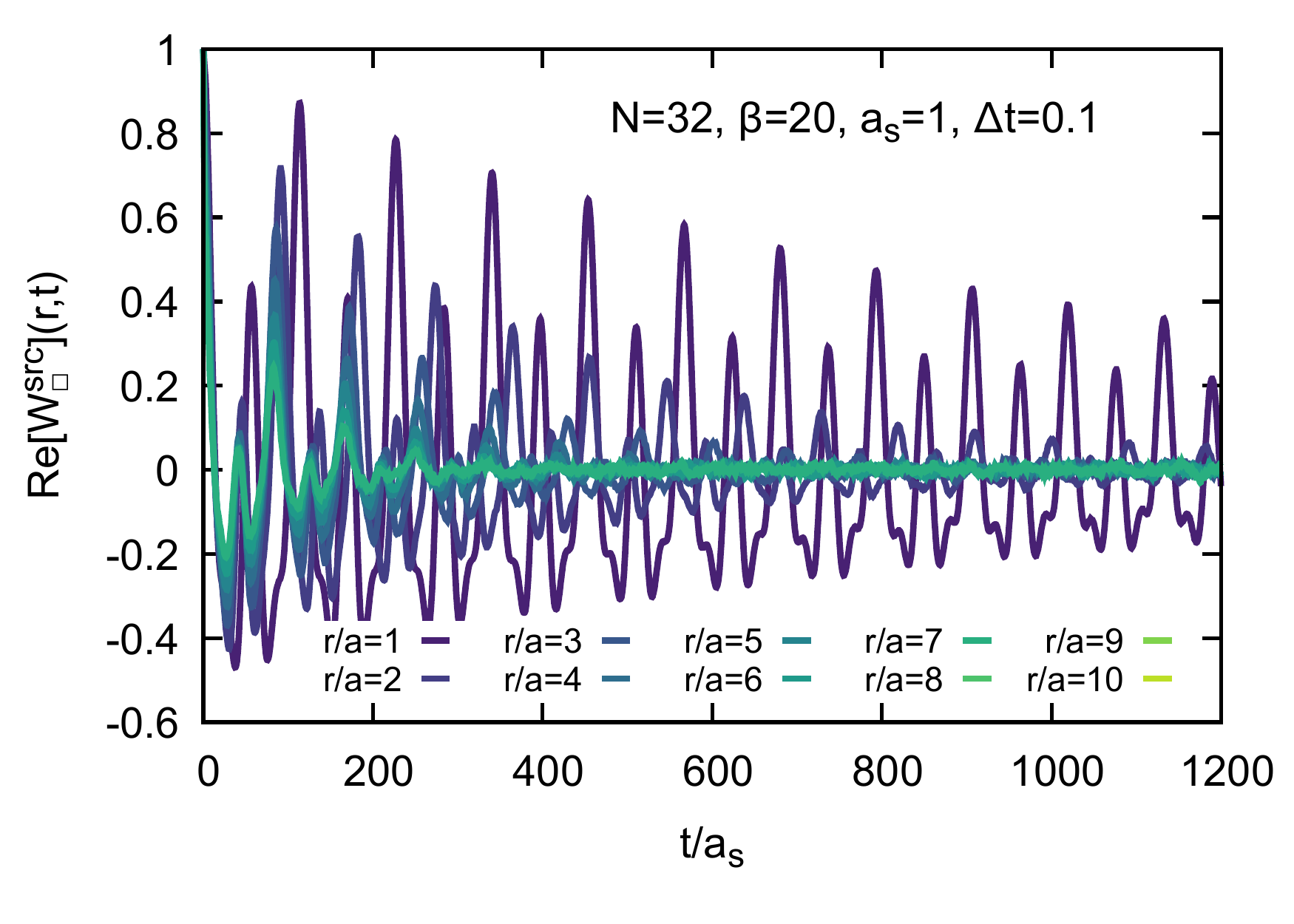}
\includegraphics[scale=0.35]{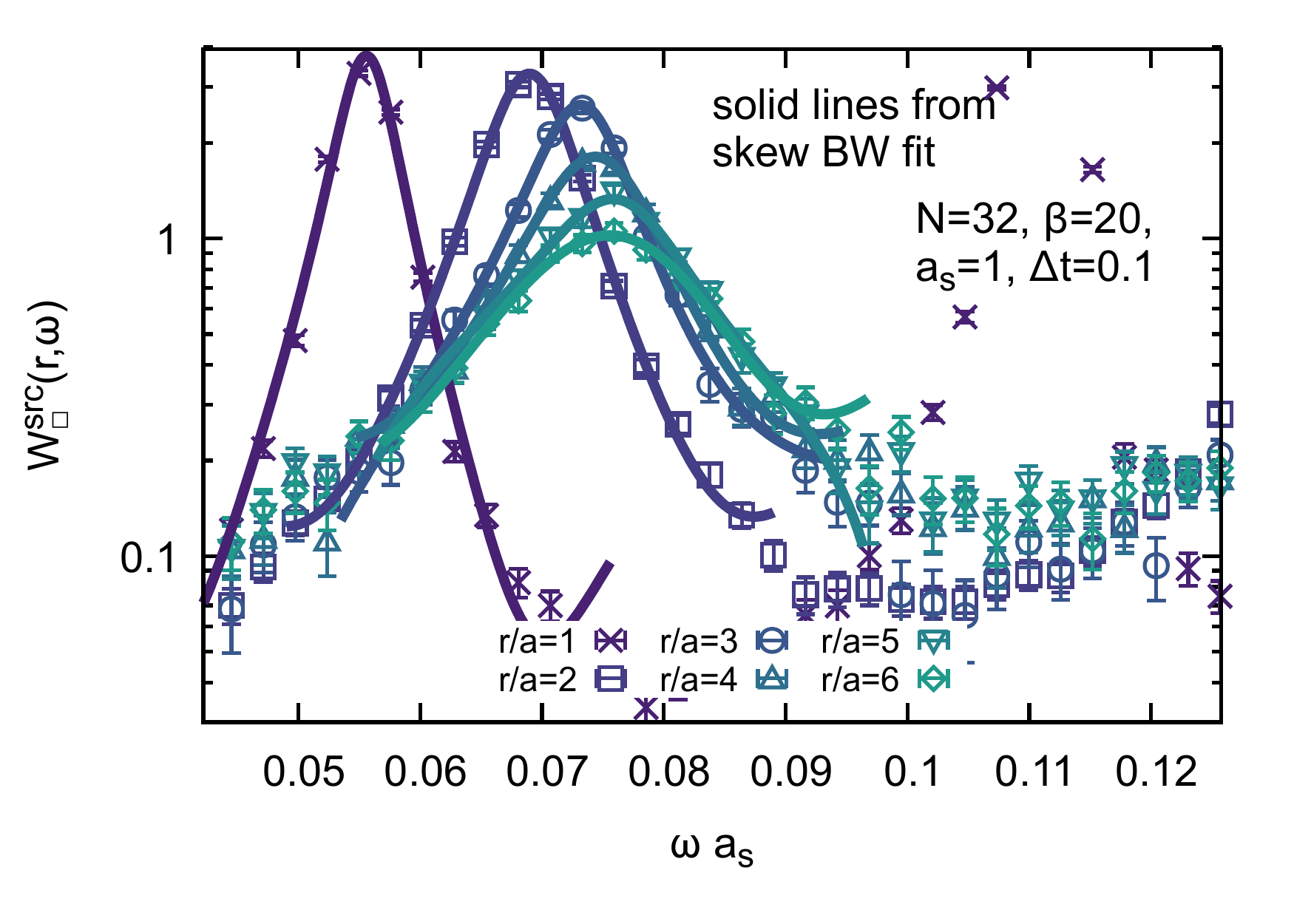}
\caption{(left) Real part of the real-time Wilson loop in CLGT in the presence of the proper Gauss law. Note the both oscillatory and damped component of the evolution. Its imaginary part shows a similar oscillatory pattern. (right) The spectral function of the real-time Wilson loop, which exhibits a well-defined low-lying peak of skewed-Lorentzian shape. Figures from \cite{Lehmann:2020fjt}.}\label{fig:CLGTWL}\vspace{-0.75cm}
\end{figure}

Solving the Yang-Mills equations of motion in the presence of the proper Gauss law significantly modifies the behavior of the Wilson loop in CLGT. In previous studies, where the charge density was absent in Gauss' law, the Wilson loop remained purely real and exhibited a simple exponential falloff at late real-times. As shown in the left panel of \cref{fig:CLGTWL} in the presence of the proper Gauss law, the real part of $\langle W_\square(r,t)\rangle$ shows an intricate oscillatory behavior with a continuously damped amplitude over time.The different colored curves correspond to different spatial separation distances, the further apart the sources the faster the damping.
Inspection of the imaginary part of $\langle W_\square(r,t)\rangle$ (see \cite{Lehmann:2020fjt}) further reveals that it differs from zero and it too exhibits damped oscillations. When one takes the (windowed) Fourier transform of the Wilson loop at different spatial separation distances, the spectral functions in the right panel of \cref{fig:CLGTWL} ensue. We see that the spectral functions exhibit a well defined lowest-lying peak structure, whose peak position and width change with the spatial separation of the static sources. Interpreted in the context of the complex interquark potential this result tells us that a finite real-part of the potential exists (finite position of the peak), as well as a finite imaginary part is present (finite width of the peak). 

\begin{figure}
\includegraphics[scale=0.35]{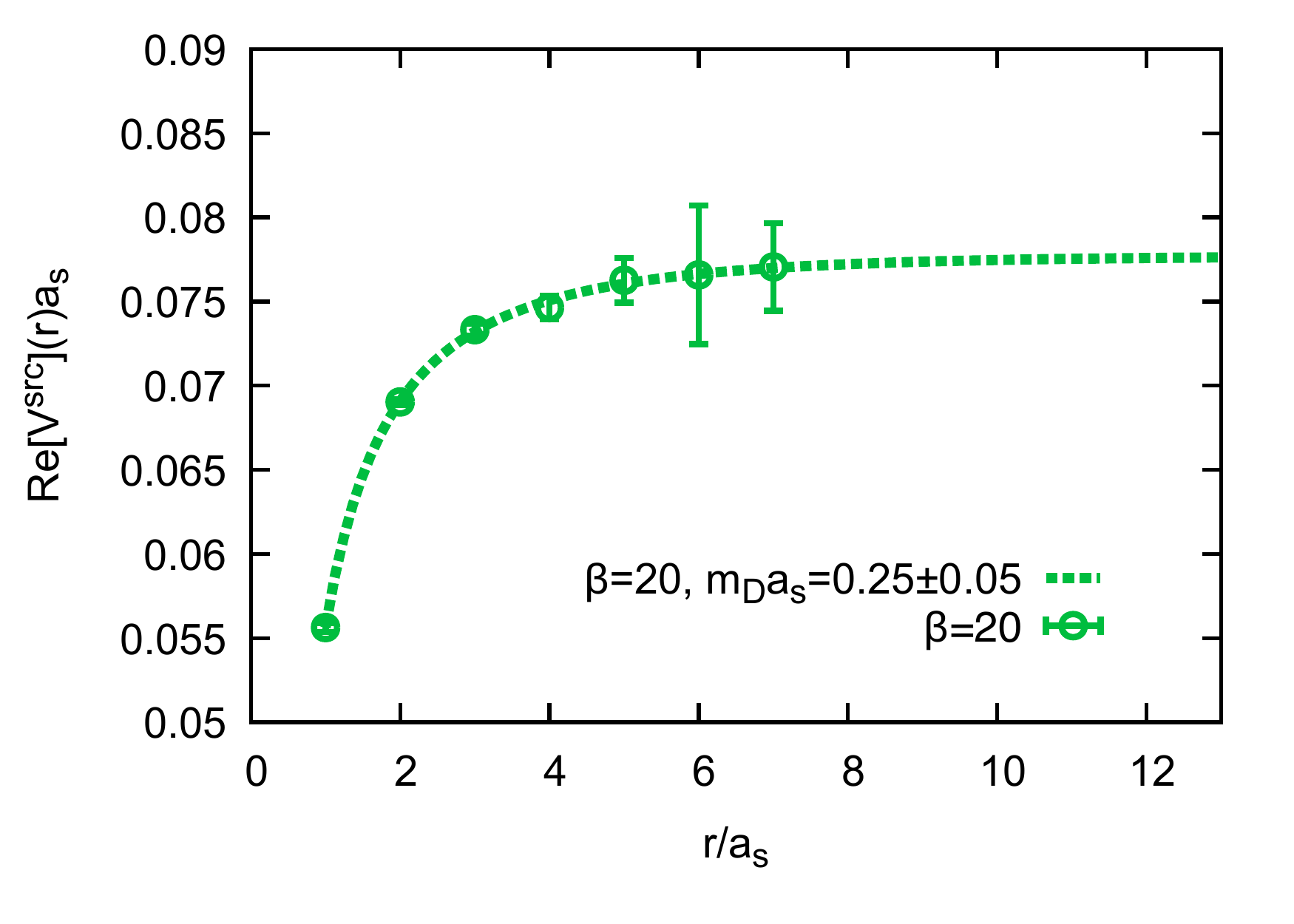}
\includegraphics[scale=0.78]{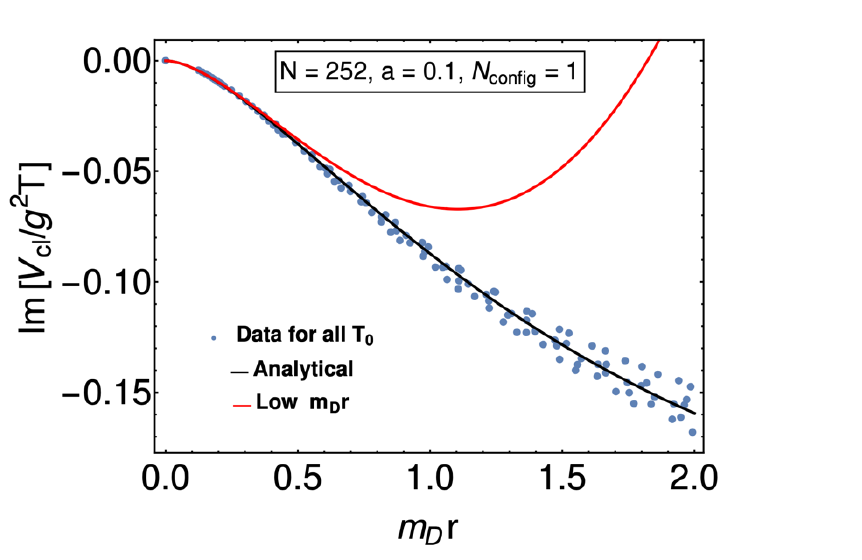}
\caption{(left) Real part of the static interquark potential in thermal classical lattice gauge theory \cite{Lehmann:2020fjt}, which is compatible with the expected Debye screening behavior in a gluonic plasma at high temperatures. (right) A high precision analysis of the imaginary part of the interquark potential in CLGT on large lattices \cite{Boguslavski:2021zga} and a comparison to a fit motivated by analytic predictions from classical hard thermal loops.}\label{fig:CLGTPot}\vspace{-0.75cm}
\end{figure}

Plotted against separation distance, we show in the left plot of \cref{fig:CLGTPot} the extracted values of the real-part of the static potential in CLGT. This is the first time that this quantity has been determined in classical statistical lattice simulations. We find that its values are compatible with an asymptotically flat form ${\rm Re}[V](r)\sim-{\rm exp}[-m_D r]/r+c$, which is reminiscent of the expected Debye screening behavior for a gluon gas at high temperature. On the left we show a recent high precision determination of the imaginary part of the static potential from a study by a collaboration between colleagues at Kent State University and TU Vienna \cite{Boguslavski:2021zga}. The large lattices deployed allow the authors to explore small enough distances to match the simulations to continuum computations of the potential from classical hard-thermal loops.

\begin{figure}
\centering
\includegraphics[scale=0.2]{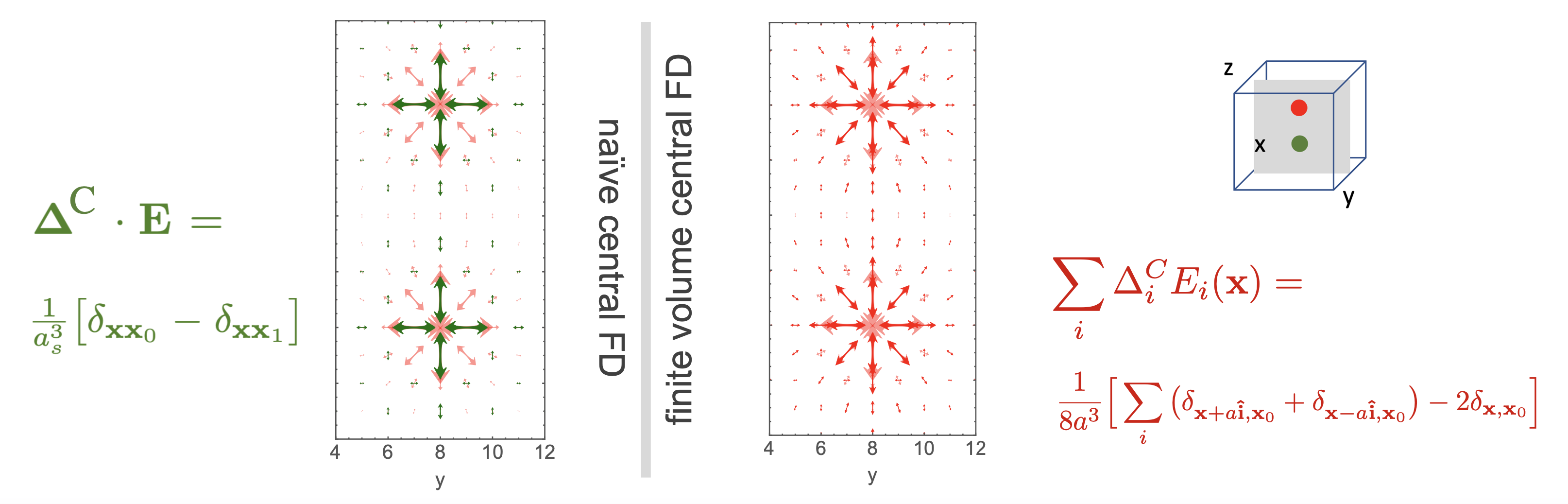}
\caption{Example of the difficulty encountered in discretizing Gauss' law with central finite differences. (left) Using a naive central finite difference discretization an artificial staggered pattern of field lines (green arrows) emerges which does not accurately reproduce the correct solution (light red arrows). (right) A finite volume discretization based on distributed sources achieves closed field lines (dark red arrows) and conserves the integral form of Gauss' law within the accuracy of the discretization.}\label{fig:GaussLaw}\vspace{-0.75cm}\end{figure}

The relevance of a correct treatment of Gauss' law on the lattice hints at that care must be taken when designing improved actions applicable to systems where translation invariance is broken e.g. by the presence of static charges. In \cite{Rothkopf:2021jye} I discuss some of the insight the computational classical electrodynamics community has gathered over the past decades. There it is recognized that two concepts are central to establish accurate finite difference based discretizations schemes. One is the use of so-called summation by parts finite difference operators (see also \cite{Rothkopf:2022zfb}), which mimic integration by parts in the discrete setting. In addition it is known that a higher order discretization of Gauss' law with a symmetric central finite difference operators requires a so-called finite volume scheme, which realizes not only the differential but also the integral form of Gauss' law. Expressed in the language of finite differences the finite-volume scheme amounts to the introduction of distributed sources, revealing an interesting relation between the concept of smeared sources and the underlying discretization of the field theory. In \cref{fig:GaussLaw} the failure of the naive central finite difference scheme for the Abelian Gauss law is shown in the left panel, where in the presence of a charge and anti-charge an artificial staggered pattern of field line emerges (green arrows). Only after going over to a finite volume scheme on the right, corresponding to a distributed charge, do we obtain closed field lines (dark red) that accurately reproduce the correct solution (light red arrows) and which satisfy the integral form of Gauss' law within the accuracy of the approximation.

Last but not least let me discuss recent developments in the direct simulation of real-time dynamics based on the complex Langevin framework \cite{namiki_stochastic_1992}. One sets out to construct a stochastic process in an artificial additional Langevin time dimension $\tau_L$ (not related to the Euclidean time $\tau$), in which the real- and imaginary part of complexified field degrees of freedom $\phi=\phi^R+i\phi^I$ are evolved in the presence of Gaussian noise $\eta$
\begin{align}
\frac{d\phi^R}{d\tau_L}={\rm Re}\left[ \left. i\frac{\delta S[\phi]}{\delta \phi}\right|_{\phi=\phi^R+i\phi^I}\right]+\eta(\tau_L),\quad \frac{d\phi^I}{d\tau_L}={\rm Im}\left[ \left. i\frac{\delta S[\phi]}{\delta \phi}\right|_{\phi=\phi^R+i\phi^I}\right].
\end{align}
The goal is to produce in the late time limit a distribution $P[\phi^R,\phi^I]$ that, when used to evaluate expectation values, reproduces the results of the Minkowski time path integral
\begin{align}
\int d\phi^R\int d\phi^I \, P[\phi^R,\phi^I]\, {\cal O}[\phi^R+i\phi^I] \leftrightarrow \int {\cal D}\phi \, {\cal O}[\phi]\, {\rm exp}[iS].
\end{align}
Traditionally there exist two major challenges: the occurrence of so-called runaway trajectories, where the stochastic process spontaneously diverges and the more fundamental limitation of convergence to incorrect results. 

Recently my group has made progress with respect to runaways, by introducing inherently stable implicit solvers to the simulation of complex Langevin \cite{Alvestad:2021hsi}. While previously runaways were combated by the introduction of an adaptive step size into explicit solvers \cite{AartsJames2010}, the implicit schemes offer additional benefits beyond inherent stability. It turns out that they also provide an inherent regularization of the underlying path integral. This has allowed us for the first time to simulate directly on the actual real-time Schwinger-Keldysh contour without the need to introduce a tilt, which so far has prevented access to phenomenologically relevant correlation functions involving the degrees of freedom on the backward contour.

At this conference we will present our most recent work \cite{Alvestad:2022hsi} that attacks the remaining challenge of complex Langevin, the possibility to converge to the wrong solution. In \texttt{Track G} Daniel Alvestad will present our novel strategy that exploits the freedom to introduce kernels into the complex Langevin equation to restore correct convergence. Our approach acknowledges that since the sign-problem is NP-hard \cite{Troyer:2004ge}, no general solution strategy is likely to succeed and therefore system specific prior knowledge needs to be incorporated into the simulation. We do so by devising a learning strategy that finds optimal kernels based on prior knowledge, such as the symmetries of the system and correlation functions accessible in Euclidean time. Our strategy allows us to extend threefold the real-time extend in which complex Langevin simulations correctly reproduce the physics of the community benchmark system, the strongly coupled anharmonic oscillator \cite{BergesSexty2007}. 

The TU-Vienna group will also present work in progress \cite{Boguslavski:2022vjz} on the simulation of real-time gauge theory on the lattice. They apply the gauge cooling and dynamic stabilization strategy developed in the context of complex Langevin for QCD at finite Baryo-chemical potential to simulations on the tilted Schwinger-Keldysh contour. For a recent proposal on how to improve real-time simulations based on the Lefschetz thimble approach see \cite{Woodward:2022pet}.

\section{Conclusion}

The determination of the real-time dynamics of strongly interacting matter lies at the heart of many open questions discussed at this conference. Lattice QCD simulations in Euclidean time have proven to be a robust tool for the study of the static properties of the strong interactions but the ill-posed inverse problem of extracting spectral functions severely limits our access to real-time dynamics. The limited information content of the Euclidean correlators presents a central challenge for conventional approaches such as Bayesian inference or more recent machine learning strategies. Incorporation of more specific QCD based prior information in the regularization of the inverse problem is thus called for.

Direct real-time simulations offer an alternative route towards gaining insight. Classical statistical lattice gauge theory neglects quantum fluctuations and may provide some qualitative insight into e.g. the binding properties of hadrons at very high temperatures. Complex Langevin simulations are steadily improving and interest in their application to the real-time setting has recently increased. Progress has been made to avoid runaways apriori using implicit solvers and new strategies to overcome the convergence to incorrect solution are actively explored.

\section{Acknowledgments}
A.R. is supported by the Research Council of Norway under the FRIPRO Young Research Talent grant 286883.

\bibliography{references}

\end{document}